\documentclass[graybox]{svmult}

\usepackage{mathptmx}       % selects Times Roman as basic font
\usepackage{helvet}         % selects Helvetica as sans-serif font
\usepackage{courier}        % selects Courier as typewriter font
\usepackage{type1cm}        % activate if the above 3 fonts are
\usepackage{makeidx}         % allows index generation
\usepackage{graphicx}        % standard LaTeX graphics tool
\usepackage{multicol}        % used for the two-column index
\usepackage[bottom]{footmisc}% places footnotes at page bottom

\usepackage{upgreek}

\usepackage{bm}
\usepackage{amsmath}
\usepackage{listings}
\usepackage{amssymb}
\usepackage{braket}
\usepackage{mathtools}
\usepackage{hyperref}
\usepackage{algorithm}

\usepackage{eurosym}
\usepackage{amstext} 
\newtheorem{dedn}{Deduction}

\usepackage[noend]{algpseudocode}

\makeindex             % used for the subject index
\begin{document}

\title*{Optimizing Execution Cost Using Stochastic Control}
\author{Akshay Bansal and Dr. Diganta Mukherjee}
\institute{Akshay Bansal \at Indian Statistical Institute, Kolkata, \email{akshaybansal14@gmail.com}
\and Dr. Diganta Mukherjee \at Indian Statistical Institute, Kolkata \email{diganta@isical.ac.in}}
\maketitle

\abstract{We devise an optimal allocation strategy for the execution of a predefined number of stocks in a given time frame using the technique of discrete-time Stochastic Control Theory for a defined market model. This market structure allows an instant execution of the \textit{market} orders and has been analyzed based on the assumption of discretized geometric movement of the stock prices. We consider two different cost functions where the first function involves just the fiscal cost while the cost function of the second kind incorporates the risks of non-strategic constrained investments along with fiscal costs. Precisely, the strategic development of constrained execution of $K$ stocks within a stipulated time frame of $T$ units is established mathematically using a well-defined stochastic behaviour of stock prices and the same is compared with some of the commonly-used execution strategies using the historical stock price data.}

\section{Introduction}
The problem of cost-efficient execution of a given stock with a lesser known distribution of its price is highly correlated with the fundamental difficulty of forecasting stock prices. A practical solution to any one of these two would bring some insight to solve the other. Investors and professional analysts frequently try to model stock prices with the help of the available information and certain noise factors whose distribution depends on various market aspects such as inflationary rates, financial status of the company and its competitive workforce. We attempt this exercise below. The rest of this section presents a formulation of the problem at hand. Section 2 discusses the execution strategy. The optimal strategy derivations are detailed in section 3. This section also discusses the algorithm for the methodology and numerical results. Finally section 4 concludes.

\subsection{Problem Formulation}\label{prob:main}

Mathematically, the execution problem can be reformulated as follows:

Determine cost-efficient policy
\begin{equation*}
\boldmath{\pi^*} = \{ \mu^*_0(x_0, R_0), \mu^*_1(x_1, R_1) \ldots \mu^*_{N}(x_{N}, R_{N})\}
\end{equation*}
such that
\begin{equation*}
\begin{aligned}
x_{k+1} &= g(x_k, u_k, \epsilon_k) \,  \forall k \in \mathbb{Z}\\
u_k &= \mu^*_k(x_k, R_k) \, \forall k \in \{0,1,\ldots,N\}\\
\sum_{r=0}^{N} u_r &= K \\  
\end{aligned}
\end{equation*}
where $g(x,u,\epsilon)$ is a known function which updates itself at each of the $N$ equispaced time points in the time duration $T$, $R_k$ is the stock position held at time point $t_k$ and $x_k$ is the stock price at time $t_k$.

Bertsimas \& Lo\cite{bertsimas1998optimal} devised one such policy by partitioning the entire time frame into $N$ intervals of equal length and performing the transaction of buying $K/N$ shares at the start of each interval. In order to analyze the expected investment cost of such policy, Bertsimas utilized the discrete form of Arithmetic Brownian Motion (ABM) ($x_t = x_{t-1} + h(u_t) + \eta \epsilon_t$) to periodically update the stock price. The major drawback with ABM model for stock price updation is that the non-negative behavior of stock price prevails only for shorter time frames T and the resultant optimal action (no. of shares bought out of the remaining stock pool) at each transaction point remained independent of any current/previous state information. Almgren \& Chriss\cite{almgren} extended the Bertismas’ model for limit order markets by incorporating the variance associated with the execution shortfall in the objective function. More recently, application of some of data-driven statistical techniques based on Reinforcement Learning\cite{suttonbarto} by Kakade et al.\cite{darkakade} and Nevmyvaka et al.\cite{kearns} have resulted in significant improvement over simpler execution strategies such as submit and leave. In 2014, Cont \& Kukanov\cite{cont} developed a more generalized mathematical framework for optimal order execution in limit order markets by incorporating targeted execution size due to bounded execution capacity of limit orders.

\section{Defining Cost-Efficient Execution Strategy}\label{sec:execute_strategy}
The uncertainty factor ($\epsilon$) involved in the state-updation function of stock price leads us to one such pathway of determining a cost-efficient policy (satisfying the conditions of (\ref{prob:main})) by minimizing the expected future cost leading to the application of well-established theory of Stochastic Control. Mathematically, the exact optimization problem reduces to determining optimal policy $\boldmath{\pi^*} = \{ \mu^*_0(x_0, R_0), \mu^*_1(x_1, R_1)$ $\ldots \mu^*_{N}(x_{N}, R_{N})\}$ for the objective
\begin{equation}\label{objective_function}
\min_{\bm\{\pi\}} \, \mathbb{E}_0[\sum_{r=0}^{N} u_r x_r]
\end{equation}
Subject to the conditions:
\begin{equation}\label{objective_constraints}
\begin{aligned}
u_k &= \mu_k(x_k, R_k) \, \forall k \in \{0,1,\ldots,N\}\\
R_{k+1} &= R_k - u_k \\
x_{k+1} &= g(x_k,u_k,\epsilon_k)\forall k\\
\sum_{r=0}^{N}u_r &= K \\
u_k &\geq 0 \, \, \forall k 
\end{aligned}
\end{equation}
where $x_k$ is the stock price at time point $t_k$, $R_k$ is the stock position held at time $t_k$ and $u_k$ is the appropriate action (investment strategy).

\begin{subsection}{Reduction to Finite horizon problem for integral states}\label{sec:finite_horizon}

The optimal policy for the problem formulated in (\ref{sec:execute_strategy}) is devised by reducing it to a finite horizon problem where the discrete time-investment is a many-to-one mapping from the tuple of stock price and remaining stocks to the countable and finite set of non-negative integers. The proposed solution to (\ref{sec:execute_strategy}) is described as following:

Given a uniform partition $\Pi(T) = \set{t_0, t_1, \ldots, t_N}$ with $X$ being the finite set of all possible stock prices and $P = \set{r \in \mathbb{Z}^{+} \mid r \leq K}$ the set of all possible stock positions. Then at any given time point $t \in \Pi(T)$, the state vector $(x,R) \in X\times P$. If the function $f(x,u,R)$ computes the \textit{instantaneous} cost for the current state $(x,R)$ and action $u$, the optimal policy ($\boldmath{\pi^*} = \{ \mu^*_0(x_0, R_0), \mu^*_1(x_1, R_1) \ldots \mu^*_{N}(x_{N}, R_{N})\}$) for the objective function (\ref{objective_function}) can be computed dynamically for each discrete time point using Bellman's principle of optimality\cite{bertsekas}. Precisely, to determine the time $t_k$ policy function $\mu^*_{k}(x_k,R_k)$, optimal action $u_k^{opt}$ is tabulated as a function of all $(x_k,R_k) \in X\times P$ using the \textit{adaptive} cost objective 
\begin{equation}\label{discrete_execution_main}
J_k(x_k,R_k) = \min_{\{u_k\}} \, \sum_{i}^{\infty} \Pr(\epsilon^i_k|\mathcal{F}_k)[f(x_k,u_k, R_k)+J_{k+1}(g(x_k,u_k,\epsilon^i_k),R_k-u_k)]
\end{equation}
where $\mathcal{F}_k$ is the $t_k$-filtration (information contained till time $t_k$).

At the final time point $t_N$, the optimal action would be to buy all the remaining $R_N$. Thus $J_N(x_N, R_N)$ simply reduces to $f(x_N, R_N, R_N)$.

If the uncertainty parameter ($\epsilon_k$) is independent of information $\mathcal{F}_k$, then (\ref{discrete_execution_main}) further simplifies to
\begin{equation}\label{discrete_execution_simplified}
J_k(x_k,R_k) = \min_{\{u_k\}} \, \sum_{i}^{\infty} \Pr(\epsilon^i_k)[f(x_k,u_k, R_k)+J_{k+1}(g(x_k,u_k,\epsilon^i_k),R_k-u_k)]
\end{equation}
At any time point $t_k \in \Pi(T)$, the optimal action $u^{opt}_k$ and $J_k(x_k,R_k)$ can be dynamically computed using (\ref{discrete_execution_simplified}) for each of the state element $(x_k, R_k) \in X \times P$.

\begin{subsubsection}{Pitfalls of reduction to integral finite horizon case}
\begin{enumerate}
	\item The numerical algorithm for its implementation mandates the construction of a three-dimensional matrix where each two-dimensional sub-matrix corresponds to a unique time point. Therefore its space complexity is of the order $\Omega(x_{max}KN)$.
	\item The method imposes an additional restraint of the finiteness and countability of the set of all possible states ($X \times P$).
	\item The numerical search for the optimal integral solution can at best be accomplished using branch and bound algorithm\cite{land_doig} whose worst case complexity is still $K$ (initial stock position). Thus the eventual time complexity for this algorithm is $\Omega(x_{max}K^2N)$.   
\end{enumerate}
\end{subsubsection}     
    
\end{subsection}

\section{Optimal Investment Strategy for instantaneous stock execution}
In this section we will develop an investment strategy based on the idea of Stochastic Control Theory discussed in \ref{sec:finite_horizon} for the market structure which sanctions the investor to buy any number of stocks on an instant basis at the current \textit{market price} (mid-point of bid-ask spread). Unlike Almgren \& Chriss\cite{almgren}, we have modelled stock prices using discretized Geometric Motion as the Bachelier's model ($x_t = x_{t-1} + h(u_t) + \eta \epsilon_t$) would eventually return negative stock prices with non-zero probability in the limit of longer time duration. The discrete time stock price model we intend to use in our analysis is given by:
\begin{equation}\label{eq:GBMstock}
x_{k+1} = x_k(1 + \beta u_k + \epsilon_k)
\end{equation}     
here $x_{t+1}$ is the stock price at time $t_{k+1}$, $\epsilon_k$ is a random noise with $\mathbb{E}[\epsilon_t]=0$ and $\beta u_k$ is the drift in stock price due to the buying action of $u_k$ no. of stocks with $\beta$ being some kind of \textit{prominence factor} which varies according to one's influence in the stock market. For our case, we'll assume $\beta$ belonging to the range $[10^{-5}, 10^{-4}]$. In the following subsections, we'll establish the general nature of some of the investment strategies for different kinds of \textit{instantaneous} cost functions and compare their performance with some well-established policies.

\subsection{Allocation Policy for Fiscal Cost Function} 
For this particular case, the \textit{instantaneous} cost function is exclusively monetary i.e $f(x_k,u_k,R_k)$ (as in section \ref{sec:finite_horizon}) is simply given by 
\begin{equation}
f(x_k,u_k,R_k) = x_k u_k 
\end{equation}
where $u_k \leq R_k$. 

Accordingly, the expression for optimal expected cost (\ref{discrete_execution_main}) modifies to
\begin{equation}\label{eq:fiscalobj}
J_k(x_k,R_k) = \min_{\{u_k\}} \, [x_k u_k + \sum_{i}^{\infty} Pr(\epsilon^i_k) J_{k+1}(g(x_k,u_k,\epsilon^i_k),R_k-u_k)]
\end{equation}

On rewriting the above expression for penultimate time point ($t = t_{N-1}$) by modelling the stock price using \ref{eq:GBMstock}, the objective simplifies to
\begin{equation}\label{eq:penobjective}
\begin{aligned}
J_{N-1}(x_{N-1},R_{N-1}) = {} & \min_{\{u_{N-1}\}} \, [x_{N-1} u_{N-1} + x_{N-1}(R_{N-1} - u_{N-1})(1 + \beta u_{N-1})] \\ 
& \text{( $\because \mathbb{E}[\epsilon_{N-1}] = 0, J_{N}(x_{N}, R_{N}) = x_{N} R_{N}$)}
\end{aligned}	 
\end{equation}
leading to the following deduction.

\begin{dedn}\label{dedn:moncost}
When the nature of the \textit{instantaneous} cost function is completely fiscal i.e. $f(x,u,R) = xu$ and the stock price is modeled using (\ref{eq:GBMstock}), the optimal investment policy due to stochastic control (Problem~\ref{objective_function})  simply converges to the purchase of the entire stock block of size $K$ at time $ t = t_N $. In general, the result holds for any stock price updation function of the form $x_{t+1} = x_t(1 + h(u_t) + \epsilon_t)$ where $h(u_t)$ is a non-decreasing drift with $h(0) = 0$	   
\end{dedn}

\begin{proof}
On rearranging the terms of penultimate time objective for the drift $h(u)$, (\ref{eq:penobjective}) modifies to 
\begin{equation}\label{obj:penultmon}
J_{N-1}(x_{N-1},R_{N-1}) = \min_{\{u_{N-1}\}} [x_{N-1}R_{N-1} + x_{N-1}(R_{N-1}-u_{N-1})h(u_{N-1})] 
\end{equation} 
As $(R_{N-1}-u_{N-1})h(u_{N-1}) \geq 0$, the optimal action ($u_{N-1}^{opt}$) results in zero with \\ $ J_{N-1}(x_{N-1},R_{N-1}) = x_{N-1} R_{N-1} $. By recursively calculating $u_k^{opt}$ and $J_k(x_k,R_k)$ using the functional form of $J_{k+1}(x_{k+1}, R_{k+1})$ (\ref{eq:fiscalobj}), it's trivial to observe the identical nature of the objective function for all $0 \leq k \leq N-1$. Hence the above deduction follows.       
\end{proof}

\begin{subsubsection}{Resultant policy and its comparison with Bertsimas' model}
Deduction~\ref{dedn:moncost} can be further generalized by observing the degenerate nature of the objective function at the penultimate time point i.e. both $0$ and $R_{N-1}$ are the optimal solutions to the objective~(\ref{obj:penultmon}). Henceforth, the optimal allocation policy modifies to the total investment for the entire stock block ($K$) at any one of the time point $t \in \set{t_0, t_1,\ldots,t_N}$.

Tabulated below is the total expenditure resulting from Bertsimas' policy and one-time investment at the midpoint $T/2$.\footnote{As per the stock data obtained from Yahoo Finance}    
\begin{table}[h]
\centering
\begin{tabular}{|c|c|c|c|} 
 \hline
 \textbf{Stock} & \textbf{Investment Cost(B)} & \textbf{Investment Cost(OT)} & \textbf{Ratio(OT:B)}  \\ [0.5ex] 
 \hline
 GOOG & \$719770.69 & \$738000 & 1.02532  \\ 
 AAPL & \$97670.42 & \$106636.21 & 1.09179\\
 QCOM & \$48983.12 & \$48808.40 & 0.99643\\
 NVDA & \$36247.86 & \$35704.41 & 0.98500\\
 LXS.DE & \euro{39972.39} & \euro{40986.76} & 1.02537 \\ [1ex]
 \hline
\end{tabular}
\caption{Comparison of total expenditure between Bertsimas'(B) and One-Time(OT) policy based on their daily opening price spanning a total of 100 working days (Feb'16 - Jun'16)}
\label{table:fiscal}
\end{table}

As evident from the data above, the one-time investment policy may frequently fail to perform better than the distributed investment policy (due to Bertsimas).   

\end{subsubsection}

\subsection{Allocation Policy for Constrained Cost Function}\label{sec:allocation_constrained}
Due to the possibility of positive accumulation of random noise ($\epsilon_t$) over large no. of discrete time steps, the allocation policy devised in the last section has a tendency of resulting in a greater investment cost compared to the policy of distributed trading over the same no. of time steps. Thus, we've made an attempt to modify the \textit{instantaneous} cost $f(x,u,R)$ by incorporating non-negative penalty in addition to the fiscal cost if the current action ($u_k$) violates certain market specific bounds. Specifically, a pre-determined set of bounds - an upper bound (UB) and a lower bound (LB) restricts the fractional consumption ($u_k/R_k$) at every time point $t_k$. The effect of penalty imposed for the case when the fractional consumption goes below the lower bound (LB) is less pronounced at initial time points compared to the later ones as the opportunistic time window to minimize the total expenditure decreases gradually with the passage of another transaction opportunity. The non-existence of such a restriction would eventually result in the investor holding a large fraction of his initial stock position at later time points with fewer opportunities to improve his total investment cost. Similarly, by restricting the investor to buy a large fraction of his current stock position (exceeding the upper bound (UB)) at the earlier time points of the transaction window, one instructs the investor to employ a distributed investment strategy till the near end of the transaction window where this constraint is liberalized. Mathematically, these two kind of restrictions can be summarized by modifying \textit{instantaneous} cost ($f(x,u,R)$) using the logarithmic barrier resulting in the functional form: 
\begin{equation}\label{eq:penaltyinstcost}
\begin{aligned}
f(x_k, u_k, R_k) = {} & x_k u_k - x_k C_l \Big(\frac{t_k}{t_N}\Big)^{\gamma} \log \Big( 1 - \max(0,LB - \frac{u_k}{R_k})\Big) \\
& - x_k C_u \Big(\frac{t_N}{t_k}\Big)^{\gamma} \log \Big(1 - \max(0,\frac{u_k}{R_k}-UB)\Big) \\ 
& \text{$\forall k \in \set{0,1,2,\ldots,N-1}$}
\end{aligned}
\end{equation}
Here $C_l, C_u$ and $\gamma$ are positive market specific constants with $C_l \gg C_u$.              

The Bellman's criteria for optimality (\ref{discrete_execution_simplified}) can now be applied for the \textit{instantaneous} cost $f(x,u,R)$ given by (\ref{eq:penaltyinstcost}) resulting in another useful deduction.
\begin{dedn}\label{dedn:ldepend}
Let $X$ be the set of all possible stock prices and the instantaneous cost $f(x_k,u_k,R_k)$ be taken of the form given by (\ref{eq:penaltyinstcost}). Then the \textit{adaptive} cost objective ($J_k(x_k,R_k)$) given by (\ref{discrete_execution_simplified}) is linearly dependent on $x_k$ ($\,\forall x_k\in X)$
\end{dedn}
\begin{proof}
Let $P(n)$ be the proposition that the cost $J_k(x_k, R_k)$ is linearly dependent on $x_k$ $\forall k \geq n$

\textit{\textbf{Base Case}}: The objective function at penultimate time point ($t_{N-1}$) is given by 
\begin{equation*}
\begin{aligned}
J_{N-1}(x_{N-1},R_{N-1}) = {} & x_{N-1} u_{N-1} - x_{N-1} C_l \Big(\frac{t_{N-1}}{t_N}\Big)^{\gamma} \log \Big( 1 - \max(0,LB - \frac{u_{N-1}}{R_{N-1}})\Big) \\ 
& - x_{N-1} C_u \Big(\frac{t_N}{t_{N-1}}\Big)^{\gamma} \log \Big(1 - \max(0,\frac{u_{N-1}}{R_{N-1}}-UB)\Big) \\
& + x_{N-1}(1 + \beta u_{N-1})(R_{N-1} - u_{N-1})
\end{aligned}
\end{equation*}
which is evidently linearly dependent on $x_{N-1}$. Thus $P(N-1)$ holds true.

\textit{\textbf{Inductive Step}}: Let $P(k+1)$ holds true for some $k \leq N-1$. Then
\begin{equation*}
\begin{aligned}
J_k(x_k, R_k) = {} & x_k u_k - x_k C_l \Big(\frac{t_k}{t_N}\Big)^{\gamma} \log \Big( 1 - \max(0,LB - \frac{u_k}{R_k})\Big) \\ 
& - x_k C_u \Big(\frac{t_N}{t_k}\Big)^{\gamma} \log \Big(1 - \max(0,\frac{u_k}{R_k}-UB)\Big) \\
& + \mathbb{E}[J_{k+1}(x_k(1+u_k+\epsilon_k),R_k - u_k)]
\end{aligned}
\end{equation*}
From the induction hypothesis, $J_{k+1}(x_k(1+u_k+\epsilon_k),R_k - u_k)$ is linearly dependent on $x_k(1+u_k+\epsilon_k)$ thus $J_k(x_k, R_k)$ is linearly dependent on $x_k$. Hence $P(n)$ holds $\forall 0 \leq n \leq N-1$ 
\end{proof}

\begin{corollary}\label{cor:indstock}
Let $X$ be the set of all possible stock prices and the instantaneous cost be taken of the form given by (\ref{eq:penaltyinstcost}). Then the optimal action $u_k$ for the objective (\ref{discrete_execution_simplified}) is independent of $x_k$ $\forall k \in \set{0,1,\ldots,N-1}$
\end{corollary}
This computationally useful corollary follows trivially from the previous deduction.

\begin{subsubsection}{Numerical Algorithm for Policy Evaluation}

The Deduction~\ref{dedn:ldepend} (and thus Corollary~\ref{cor:indstock}) is extremely advantageous to develop an efficient algorithm for determining the policy as the optimal action resulting from the theory of stochastic control is independent of stock price $x$. Hence all future computations can be performed by assuming stock price to be \textit{unity}. 

\begin{algorithm}
\caption{An efficient algorithm to compute optimal policy for constrained cost}\label{alg:constrainedcost}
\begin{algorithmic}[1]

%\Procedure{Optimal_Allocation}{$J,U$}\Comment{2-D arrays to store optimal cost and action}
\While{$r \in \{0,1,\dots,InitialSize\}$}
\State $J[N][r] \gets r$\Comment{Optimal Cost at time $t_N$ for $x=1$}
\State $U[N][r] \gets r$\Comment{Optimal action at $t_N$ (ind. of $x$ from last corollary)}
\EndWhile

\While{$ i \in \{N-1,N-2,\dots,0\}$}\Comment{To evaluate $u_{opt}$ and $J_i(1,r)$ at each time point $t_i$}
	\State$J[i][0] \gets 0$\Comment{Optimal Cost when stock position is null}
	\State$U[i][0] \gets 0$\Comment{Optimal Action when stock position is null}
	\While{$r \in \{1,2,\dots,InitialSize\}$}\Comment{To determine the optimal action for each possible stock position}
		\State$u_{opt} \gets 0$
		\State$val_{opt} \gets f(1,u_{opt},r) + (1+\beta u_{opt})J[i+1][R-u_{opt}]$	
	
		\While{$u \in \{1,2,\dots,r\}$}\Comment{Brute force search to determine optimal action dynamically}
			\State$val_u \gets f(1,u,r) + (1 + \beta u)J[i+1][R-u] $
			\If{$val_u \leq val_{opt}$}
				\State$u_{opt} \gets u$
				\State$val_{opt} \gets val_u$
			\EndIf
		\EndWhile
		\State$J[i][r] \gets val_{opt}$
		\State$U[i][r] \gets u_{opt}$
	\EndWhile
\EndWhile

%\EndProcedure
\end{algorithmic}
\end{algorithm}

\end{subsubsection}

\begin{subsubsection}{Resultant policy for constrained objective}

The optimal allocation vector (in row-major form) with initial stock position of $1000$ shares for $\beta = 5 \times 10^{-5}$, $C_l = 1000$, $C_u = 10$, $\gamma = 2$, $LB = 0.2$, $UB = 0.6$ and different number of time points is depicted as under
\begin{itemize}
	\item $N = 10$
	\begin{equation*}
		\vec{u}^{opt} = \begin{bmatrix}
						600 & 240 & 96 & 38 & 15 & 6 & 1 & 2 & 1 & 1
						\end{bmatrix}
	\end{equation*}

	\item $N = 30$
	\begin{equation*}
		\vec{u}^{opt} = 		\begin{bmatrix}
0 & 600 & 110 & 58 & 47 & 37 & 30 & 24 & 19 & 15  \\ 
12 & 10 & 8 & 6 & 5 & 4 & 3 & 3 & 2 & 2  \\
1 & 1 & 1 & 1 & 1 & 0 & 0 & 0 & 0 & 0 
					\end{bmatrix}
	\end{equation*}

	\item $N = 50$
	\begin{equation*}
		\vec{u}^{opt} = \begin{bmatrix}
0 & 0 & 0 & 600 & 39 & 72 & 58 & 46 & 37 & 30 \\ 
24 & 19 & 15 & 12 & 10 & 8 & 6 & 5 & 4 & 3 \\ 
3 & 2 & 2 & 1 & 1 & 1 & 1 & 1 & 0 & 0 \\
0 & 0 & 0 & 0 & 0 & 0 & 0 & 0 & 0 & 0 \\ 
0 & 0 & 0 & 0 & 0 & 0 & 0 & 0 & 0 & 0 
						\end{bmatrix}
	\end{equation*}

	\item $N = 100$
	\begin{equation*}
		\vec{u}^{opt} = \begin{bmatrix}
0 & 0 & 0 & 0 & 0 & 0 & 0 & 0 & 0 & 119 \\
176 & 141 & 113 & 90 & 72 & 58 & 46 & 37 & 30 & 24 \\
19 & 15 & 12 & 10 & 8 & 6 & 5 & 4 & 3 & 3 \\ 
2 & 2 & 1 & 1 & 1 & 1 & 1 & 0 & 0 & 0 \\
0 & 0 & 0 & 0 & 0 & 0 & 0 & 0 & 0 & 0 \\
0 & 0 & 0 & 0 & 0 & 0 & 0 & 0 & 0 & 0 \\
0 & 0 & 0 & 0 & 0 & 0 & 0 & 0 & 0 & 0 \\
0 & 0 & 0 & 0 & 0 & 0 & 0 & 0 & 0 & 0 \\
0 & 0 & 0 & 0 & 0 & 0 & 0 & 0 & 0 & 0 \\ 
0 & 0 & 0 & 0 & 0 & 0 & 0 & 0 & 0 & 0 

						\end{bmatrix}
	\end{equation*}

\end{itemize}

With a steady increment in the number of available time points $N$ for transaction in the fixed interval $T$, the resultant allocation policy follows a strategy of smaller stock acquisition towards the beginning and end of the interval $T$ whereas bigger transactions are made towards the middle. Intuitively, this kind of allocation behaviour can be explained the observing the effect of the drift $\beta u$ which has a tendency to increase the stock price resulting in a larger investment cost. Therefore, it is advantageous to make small transactions towards the beginning in such a way that the stock prices have a little tendency to drift upwards and at the same time a noticeable fraction of the initial stock position is also fulfilled followed by a major acquisition towards the middle. The resultant hefty drift would eventually have a little effect on the total investment cost as the remaining stocks constitute a small fraction of the initial stock position $K$.

\end{subsubsection}

\begin{section}{Conclusion}
The policy resulting from the analysis performed in section \ref{sec:allocation_constrained} by incorporating several risk-factors has shown considerable improvement over the Bertsimas' policy with its total expenditure tabulated as under:\footnote{As per the stock data obtained from Yahoo Finance}
\begin{table}[h]
\centering
\begin{tabular}{|c|c|c|c|} 
 \hline
 \textbf{Stock} & \textbf{Investment Cost(B)} & \textbf{Investment Cost(WR)} & \textbf{Ratio(WR:B)}  \\ [0.5ex] 
 \hline
 GOOG & \$719770.69 & \$699576.13 & 0.97194\\ 
 AAPL & \$97670.42 & \$94117.02 & 0.96361\\
 QCOM & \$48983.12 & \$45666.70 & 0.93229\\
 NVDA & \$36247.86 & \$28670.91 & 0.79096\\
 LXS.DE & \euro{39972.39} & \euro{35319.80} & 0.88360\\ [1ex]
 \hline
\end{tabular}
\caption{Comparison of total expenditure between Bertsimas' (B) and Cost with Risks (WR) based on their daily opening price spanning a total of 100 working days (Feb'16 - Jun'16)}
\label{table:penaltycost}
\end{table}

In summary, the non-performance of one-time investment policy (Table~\ref{table:fiscal}) and significant improvement of the policy resulting from the modified cost function (Table~\ref{table:penaltycost}) by incorporating market risks can be safely established for the average case analysis of market model-I keeping in mind the existence of a non-zero probability of the occurrence of a case scenario where the above deduction fails to hold.

The \textit{instantaneous} cost objective~(\ref{eq:penaltyinstcost}) could be improved further by factoring constraints in a rational manner such that the penalty levied upon their violation does not undermine or overestimate the effective fiscal cost. Another way to improve the cost objective is by estimating the effect of current stock price before converging to any possible action. For instance, if the bounds on the possible stock prices and its probability distribution throughout the entire time duration $T$ is already known, then one can possibly make use of this information by tuning the penalty functions appropriately as a significantly lower stock price and higher probability density would result in a net reduced risk for the case when one intends to invest in a large fraction even at the earlier time points. Similarly, a higher price (close to upper bound) would levy a high penalty even when one is within the bounds of the imposed constraints. These kind of formulations would bring in the dependence of the stock price resulting in improved policies but with a slight trade-off of an increased time and space complexity.

Another possible way to improve the performance of the resulting control action is by utilizing a more general form of the stock price updation function based on the theory of Linear Price Impact with Information as suggested in  i.e. the stock price at each successive time point can now be modeled as:
\begin{equation}
\begin{aligned}
x_{t+1} &= f(x_t, u_t, Z_t, \epsilon_t) \\
Z_t &= g(Z_{t-1}, \eta_t)
\end{aligned}
\end{equation}

\end{section}

\begin{acknowledgement}
We thank the conference participants at Statfin2017 for their insightful comments and suggestions which has helped us to improve our work.
\end{acknowledgement}

%%%%%%%%%%%%%%%%%%%%%%%% referenc.tex %%%%%%%%%%%%%%%%%%%%%%%%%%%%%%
% sample references
% %
% Use this file as a template for your own input.
%
%%%%%%%%%%%%%%%%%%%%%%%% Springer-Verlag %%%%%%%%%%%%%%%%%%%%%%%%%%
%
% BibTeX users please use
% \bibliographystyle{}
% \bibliography{}
%

\end{document}